\documentclass[aps,floatfix,showkeys,superscriptaddress,notitlepage]{
revtex4}
\usepackage{amsmath,amssymb,amsfonts,graphics,graphicx,dcolumn,bm,enumerate}
\usepackage{comment,natbib,appendix}
\usepackage{multirow,color}

\usepackage{amsthm}
\usepackage{epsfig}
\usepackage{natbib}
\usepackage{hyperref}
\usepackage{graphicx}
\usepackage{epstopdf}

\newcommand{\sinp}
{\affiliation{Condensed Matter Physics Division, 
Saha Institute of Nuclear Physics, 1/AF Bidhannagar, Kolkata 700064, India.}}
\newcommand{\iima}
{\affiliation{Economics area, Indian Institute of Management, Vastrapur, 
Ahmedabad 380015, India}}
\newcommand{\aalto}
{\affiliation{Department of Computer Science, Aalto University School of 
Science,
P.O. Box 15400, FI-00076 AALTO, Finland}}
\newcommand{\jnu}
{\affiliation{School of Computational and Integrative Sciences, Jawaharlal 
Nehru 
University, New Delhi 110067, India}}
\newcommand{\csss}
{\affiliation{CTRPFP, Centre for Studies in Social Sciences, Calcutta, 
R–1 Baishnabghata Patuli Township, Kolkata 700094, India}}

\begin{document}

\title{Quantifying invariant features of within-group inequality in consumption 
across groups}

\author{Anindya S. Chakrabarti}
\email[Email: ]{anindyac@iimahd.ernet.in}
\iima 
\author{Arnab Chatterjee}%
\email[Email: ]{arnabchat@gmail.com}
\sinp
\author{Tushar K. Nandi}
\email[Email: ]{nandi.tushar@gmail.com}
\csss
\author{Asim Ghosh}
\email[Email: ]{asimghosh066@gmail.com}
\aalto
\author{Anirban Chakraborti}
\email[Email: ]{anirban@jnu.ac.in}
\jnu

\begin{abstract}
We study unit-level expenditure on consumption across multiple countries and 
multiple years, in order to extract
invariant features of consumption distribution. We show that the
bulk of it is lognormally distributed,
followed by a power law tail at the limit. The distributions coincide with each other under  
normalization by mean expenditure and log scaling even though the data is sampled across multiple dimension including, e.g., 
time, social structure and locations. This phenomenon indicates that the dispersions in consumption expenditure 
across various social and economic groups
are significantly similar subject to suitable scaling and normalization. 
Further, the results provide a measurement of the core distributional 
features. Other descriptive factors including those of sociological, demographic 
and political nature,
add further layers of variation on the this core distribution.
We present a stochastic multiplicative model to 
quantitatively characterize the invariance and the distributional features.
\end{abstract}

\keywords{Inequality, invariance, consumption distribution, power law, 
lognormal distribution.}


\maketitle

\textit{Any city, however small, is in fact divided into two, one the city of 
the poor, the other of the rich \ldots - Plato~\cite{Plato} }

\section{Introduction}

Plato's remark as stated at the beginning refers to the intuitive notion that even though the reason for inequality could be vastly different, but the dispersion in affluence is
always present. Seminal work by \cite{Pareto-book} shows that the right tail of the wealth distribution has a power law tail. Further explorations have
shown that this feature is invariant across countries although the origin of the power law is not settled yet, either theoretically and empirically. 
The existence of a fat tail constitutes one such invariance even though the exponent and the share accruing to the
top income classes are seen to be fluctuating substantially across countries and time (\cite{PikettyAtkinson2010}, see also \cite{chakrabarti2013econophysics}).
The bulk of the distributions are also described well by lognormal or gamma distributions (\cite{chakrabarti2013econophysics}), which again is susceptible to substantial variation in parameters even though
the functional form remains the same. Thus there is hardly any precise and specific quantitative feature which is common across samples.

Quantifying inequality has been one of the most important factors in devising 
economic policies targeted towards mitigating the same.
Theoretical tools developed for that purpose are equipped to find out the level of inequality based on a vector of income or wealth from a sample
of units. Depending on the case, the unit could be an individual or a household (or something else). Keeping track of the level of inequality for the same sample
over time or the same across different samples collected at the same point in time, allows us to make comparative judgments about the dynamics of inequality. 
In this paper, we ask the question: is there any fundamental feature of inequality that is invariant across samples (both across time and geographic boundary)? 

We argue that in case of consumptions, the mean of the distribution is an important scaling factor. Once the distributions are normalized by their respective mean values,
the inequality of the normalized sample show reasonable agreement in terms of numerical values. We study data with large sample size across three countries (India, Brazil and Italy)
and a number of years (distinct waves when the data were collected). Each country is a fiscal and monetary union of smaller states and/or other units e.g.
religious or ethnic groups. The financial markets are also more integrated within each country than across countries. Both of these imply that
the consumption decisions faced by households within a country are made in an environment much similar than households across countries.
We show that within each country, the consumption distribution across different economic or social identities (states or religions or locations)
show almost identical features once normalized by the respective mean values.
The choice of set of countries (India, Brazil and Italy) under study stems from data availability.

To account for the distributional features, we provide a small-scale heterogeneous households model to quantify the dispersion and the existence of both lognormal bulk and power law tail. 
Previous models had either focused on the bulk which is lognormally distributed e.g. the literature that builds upon the approach proposed by Kalecki or the tail, which is power law distributed (see 
\cite{chakrabarti2013econophysics}). In the present paper, we propose a mixture model that is able to generate both simultaneously. 
In the model, we assume that households' consumption decisions are affected by habit. 
They interact through a capital market and receive idiosyncratic shocks in labor income that they cannot smooth out. Such incompleteness in the market along with
heterogeneity in habits across households, generate a distribution of consumption. Using tools from distributional analysis, we show that the distribution has
a dominant power law component in the limit and a lognormal bulk.

\cite{CCGCN;16} was an initial attempt to study if there is any invariance in consumption. However, the scope of
that study was very limited due to data availability (India 66th round, year 2009-10). In the present context, we have analyzed a much bigger data set
from multiple countries spanning over multiple years.
This paper is related to two strands of literature. One, we invoke the idea of invariance in distributions of economic quantities like income or wealth
(\cite{Pareto-book}). We differentiate our work from \cite{Kuznets;55} who stressed the evolution of inequality across time due to evolution
of market institutions. In a similar vain, \cite{Acemoglu-nations-fail;13} proposed a theory of historical evolution of inequality
as a reflection of the evolution of political institutional features. A different version was proposed by \cite{Galor_book_11}which emphasized development of institutions, specially educational sector being an important factor. Our approach is complementary in that we propose that there always exists a substantial level of inequality conditional on the
state of the economy which is captured by the average affluence. On the technical side, there are multiple attempts to model the power law structure which is the most commonly known invariant feature of income/wealth distributions. \cite{chakrabarti2013econophysics}
contains a number of models in that direction.
\cite{Benhabib_14} showed that it is possible to generate a power law in the tail by using an overlapping generation framework with incomplete markets.
As we have discussed in the modeling section in details, we use the specifications in \cite{Gabaix_11} and 
\cite{Kelly_13} for analytical purpose.

In the next section, we describe the data and summary statistics of consumption distribution.  
In the following section, we present the key results regarding variation in consumption across
countries and time. Finally, to account for the robust pattern we see in data, we present a simple stochastic model
of consumption distribution.

\section{Data description}

We use the data for Household Consumer Expenditure 68th Round (2011-2012) from 
the National Sample Survey Office ~\cite{NSSO} of India.
It contains information 
about expenditure incurred by households on consumption goods and services 
during the reference period. These sample surveys are conducted 
using households as unit of the economy. 
This ignores heterogeneity in household size but the data contains 
information about monthly per capita consumer expenditure  (MPCE) in Indian Rupee (INR).
Data is available for all sampled households in the different states and 
Union territories (UT), across several parameters like castes, religions and 
rural-urban divide.
\cite{CCGCN;16} studied Household Consumer Expenditure 66th Round (2009-2010) collected by
the National Sample Survey Office (NSSO) which collected data for multiple definitions of expenditure and used multiple definitions of inequality. The important conclusions we draw from that study is
that the results are robust to such changes in definitions. So we focus on very specific and standard definitions only, in the present work.
There are 101717 households in this data set (see 
tables~\ref{tab:india68a} and \ref{tab:india68b}).
To study the inequality structure, we use two kinds of data which provides two 
perspectives.

Data from Brazil is procured from IBGE- Instituto Brasileiro de Geografia e 
Estat\'{i}stica. The Consumer Expenditure Survey data from two rounds 
(\cite{Brazil0203} with 48470 households and  
\cite{Brazil0809} with 55970 households) are used here 
(table~\ref{tab:brazil_states}). The data contains information about household 
size, geographical location (state) and consumer expenditure in Brazilian Real (BRL) for different 
expenditure sectors, among other things.

For Italy, we use microdata provided by \cite{Italydata}
and has information about household consumer expenditure in Euro (EUR).
We analyze 10 years of data (1980-1984, 1986, 1987,1989, 1991, 1993, 1995, 
1998, 2000, 2002, 2004, 2006, 2008, 2010, 2012).
See App.~\ref{sec:app_data} for description and summary statistics.

\section{Results}
In this section, we discuss the main results of data analysis. First feature is that the normalized data collapses on one single distribution.
The second feature is that the distribution is lognormal followed by power law tail.
\subsection{Invariance}
The available consumption data shows a scaling property across time and countries. Consider a variable
$x$ which denotes consumption expenditure. Suppose it has a distribution $p_{it}(x)$ in cross section, in the 
$i-$th country, $t-$th period. We show that by choosing a suitable scaling parameter,
the data collapses into one single aggregate distribution across different countries and years upon taking log transformation. That is,
the scaled variable
\begin{equation}
X_{it}=\log\left(\frac{x_{it}}{\tau^{1/\kappa}_{it}}\right)
\label{Eqn:scaling}
\end{equation}
has a distribution
\begin{equation}
p_{it}(X)=p(X)
\end{equation}
for all $i$ and $t$.
The parameters we chose, are mean consumption expenditure 
($\tau_{it}=E(x_{it})$ where $E(.)$ denotes expectation operator) and 
$\kappa=1$.
Collapse of data onto a single distribution indicates that there is a core 
inequality process which is generated and described by mechanisms similar across 
geographic boundary and time. 

\subsection{Normalized distributions}
\label{Sec:app-normalization}
Consider a variable $x$ following a lognormal distribution,
\begin{equation}
f(x)=\frac{1}{x\sigma\sqrt{2\pi}} \exp\left(- \frac{(\log x 
-\mu_x)^2}{2\sigma_x^2} \right).
\end{equation}
The mean of this distribution is
\begin{equation}
E(x)=e^{\mu_x+\frac{\sigma^2_x}{2}}
\end{equation}
and the variance is given by
\begin{equation}
V(x)=(e^{\sigma^2_x}-1).e^{2\mu_x+\sigma^2_x}.
\end{equation}
Thus upon normalization by the mean, the new distribution has a mean of
\begin{equation}
E^{norm}(x)=e^0=1
\end{equation}
and variance,
\begin{equation}
V^{norm}(x)=(e^{\sigma^2_x}-1).
\end{equation}
By linearizing the exponential term in the variance we get
\begin{equation}
V(x)=\sigma^2_x+2\mu \sigma^2_x+\sigma^4_x
\end{equation}
whereas for the normalized variable we have
\begin{equation}
V^{norm}(x)=\sigma^2_x.
\end{equation}
The tail of the distribution is found to be power law or a 
Pareto distribution which can be represented as
\begin{equation}
p(x)\sim x^{-(1+\gamma)}.
\end{equation}
By the nature of the distribution, power law is scale-free i.e. normalization
of data which is power law distributed, leaves the distributional 
features unchanged.

\begin{figure}
\includegraphics[width=17.0cm]{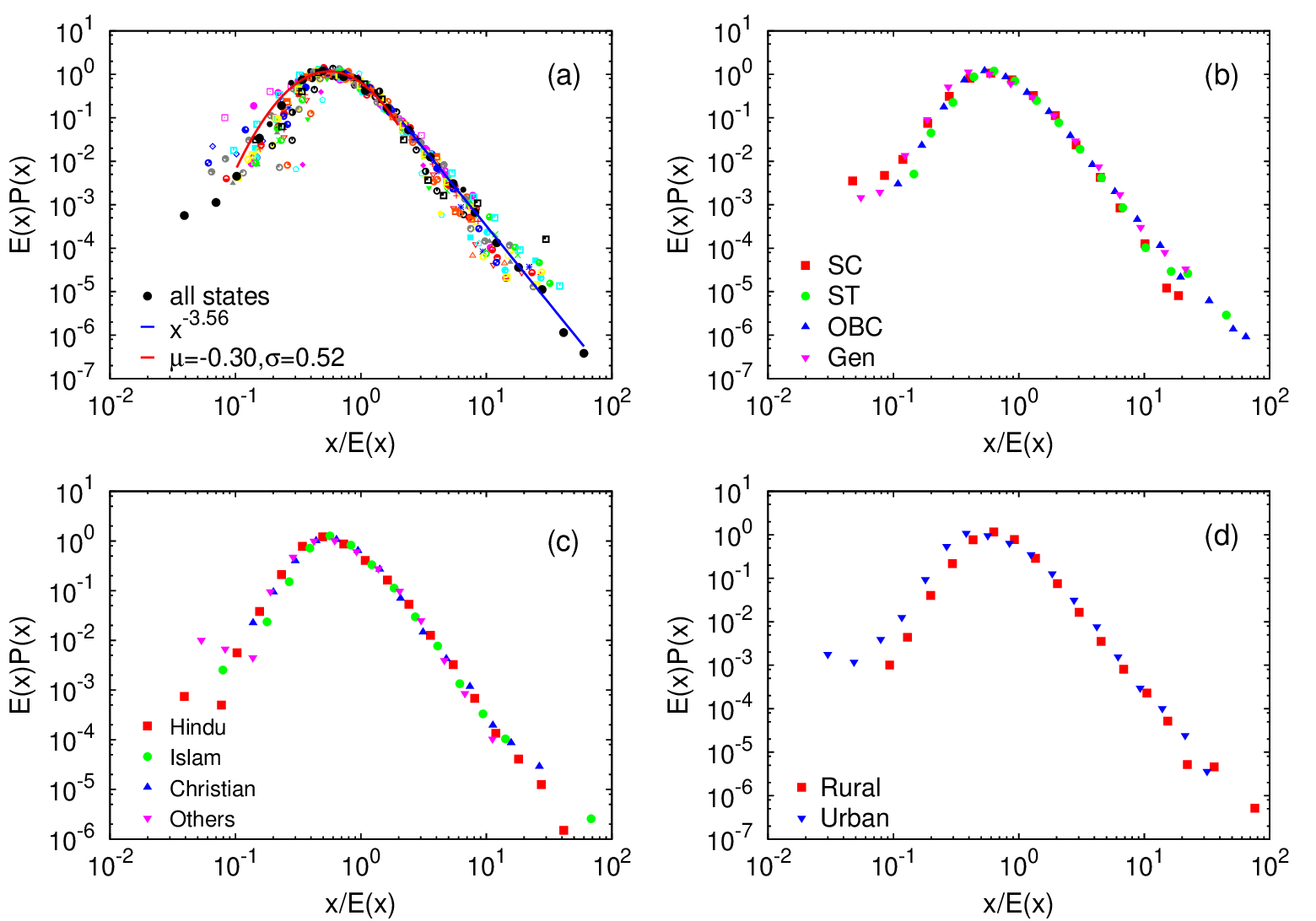}
 \caption{Collapse of consumption expenditure data for Indian states. Data collected for round 68 (2011-2012). Panel (a): Normalized data for all states.
Fitted with a lognormal distribution and a power law at the right tail.
Panel (b): Normalized data across caste categorization. 
Panel (c): Normalized data across religious categorization.
Panel (d): Normalized data across urban and rural population.}
 \label{fig:India_all_11_12}
\end{figure}

The scaled distribution in all cases show a lognormal bulk and a power law asymptotically. This form has also been argued to be extracted from income and wealth data 
(\cite{chakrabarti2013econophysics}).
The difference is that here, the exponent of the power law tail is much higher than that of income or wealth distributions indicating much faster rate of decay and
lower inequality. The finding that consumption is less dispersed than income is consistent with the available evidence (\cite{Christiano_87}). 
Finally, we note that Eqn. \ref{Eqn:scaling} is a linear transformation of the original expenditure variable and hence the distributional features remain intact. Only
the moments change due to this transformation. See Sec. \ref{Sec:app-normalization} for a short description of the parametric features of the normalized distribution.

\begin{figure}
\centering
\includegraphics[width=14.0cm]{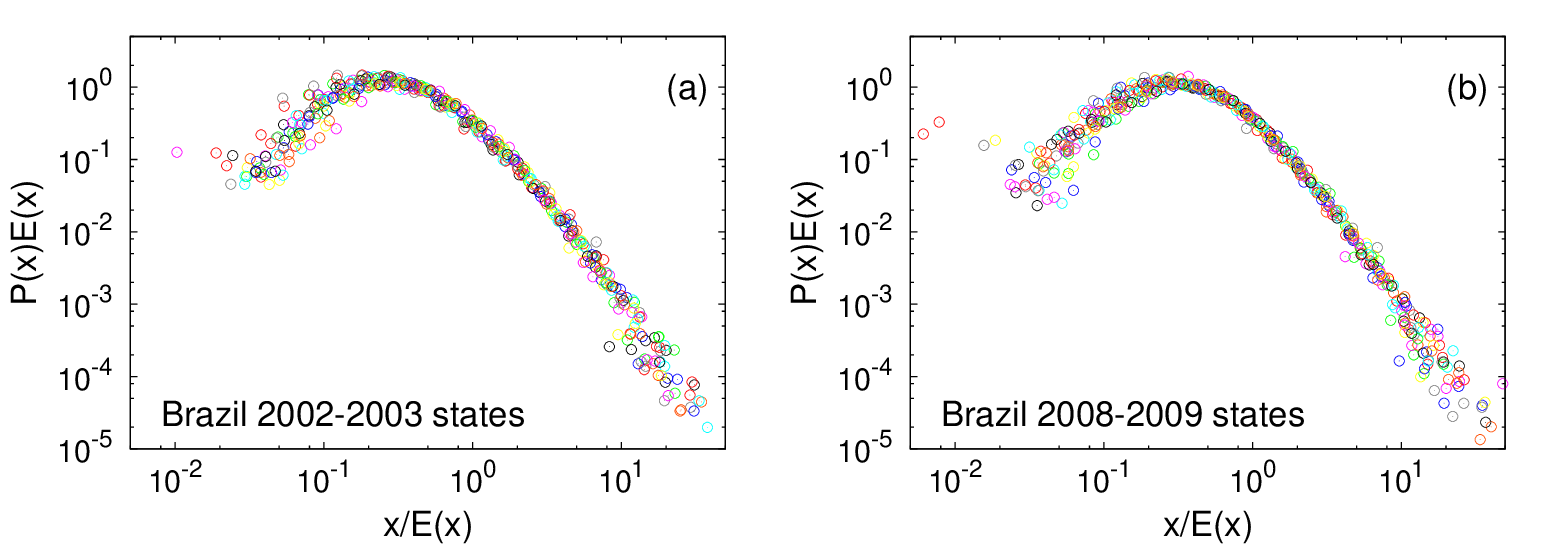}
 \caption{Consumption expenditure data across Brazilian states normalized with respect to the respective mean expenditure across states; Panel (a): year 2002-03, Panel (b): year 2008-09. See table \ref{tab:brazil_states}
for details.}
 \label{fig:Brazil_allstates}
\end{figure}
\begin{figure}
\centering
\includegraphics[width=9.0cm]{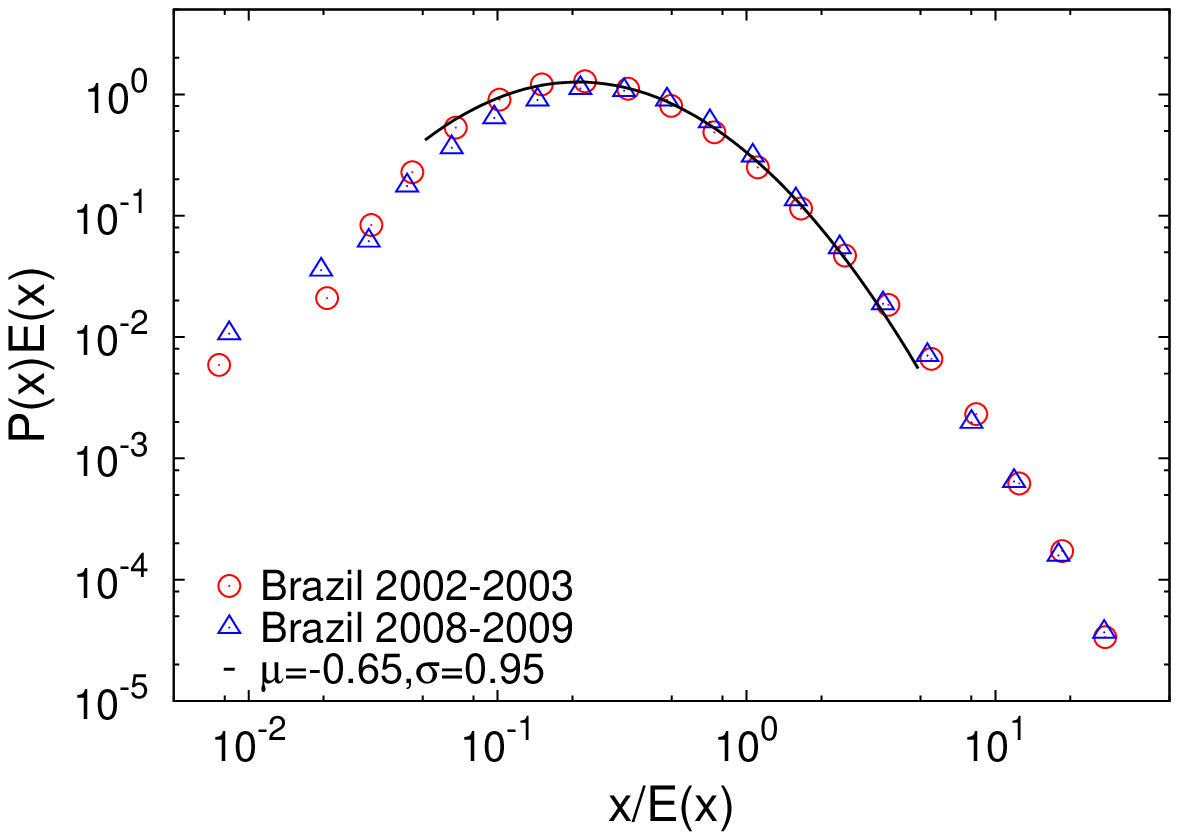}
 \caption{Consumption expenditure data across years in all states of Brazil combined. Panel (a): year 2002-03, Panel (b): year 2008-09.}
 \label{fig:Brazil_allyears}
\end{figure}
\begin{figure}
\centering
\includegraphics[width=10.0cm]{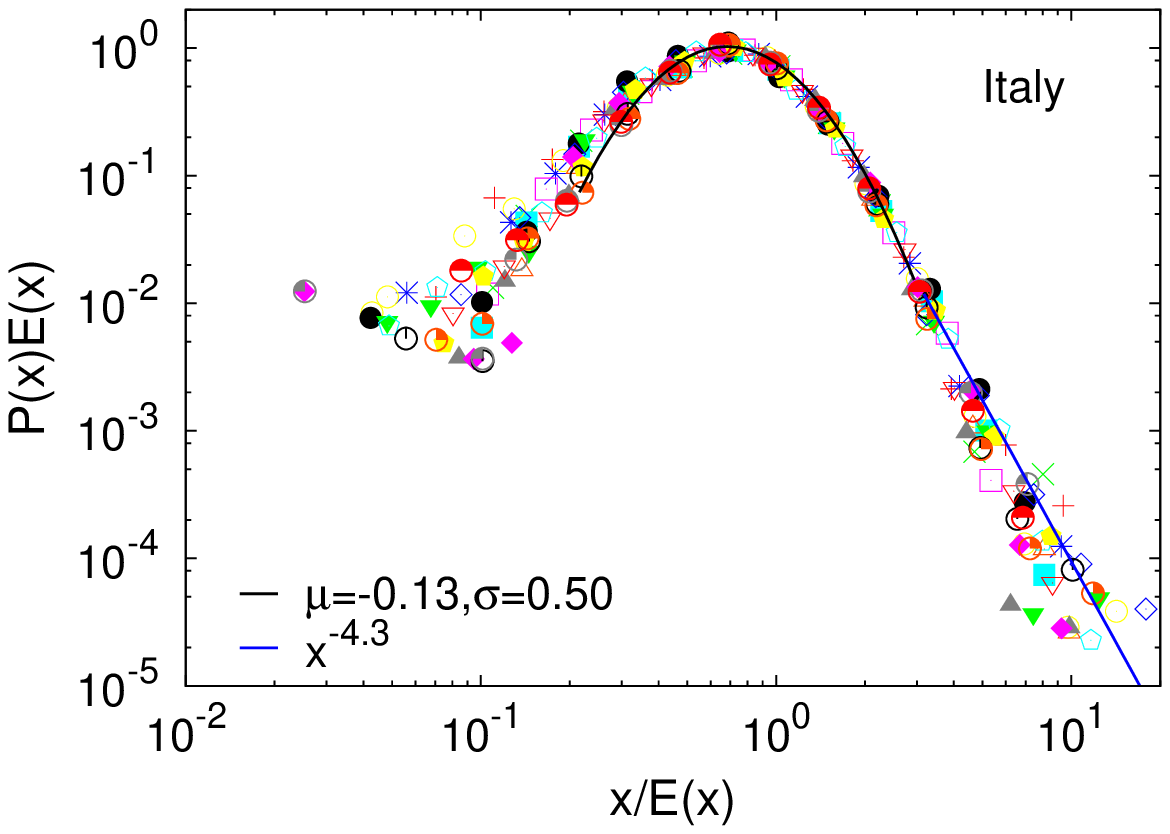}
 \caption{Consumption expenditure data across years in all states of Italy combined. Data ranges from 1980-2012 (see table 
\ref{tab:italy} in appendix).}
 \label{fig:Italy_allyears}
\end{figure}

We present the distribution of the scaled expenditure variable in 
Fig. \ref{fig:India_all_11_12} in for Indian states and also for other dimensions including
caste (panel b), religion (panel c) and urbanity (panel d). 
This figure
with superimposition of the cross-sectional data
shows that the distributions coincide under normalization which is
consistent with the preliminary findings made by \cite{CCGCN;16} for 
different wave of data collection. 
The bulk of the data fits with lognormal distribution
and the tail fits with a power law.
Fig. \ref{fig:Brazil_allstates} shows a similar data 
collapse in case of Brazil across all states in two given years
and Fig. \ref{fig:Brazil_allyears}
shows for multiple years across all states. The data is fitted with a lognormal distribution.
In Fig. \ref{fig:Italy_allyears}, we present normalized Italian data
across years. Similar to the Indian data set, the bulk fits with lognormal
distribution and the tail is fitted with a power law. 

\begin{figure}
\centering
\includegraphics[width=15.0cm]{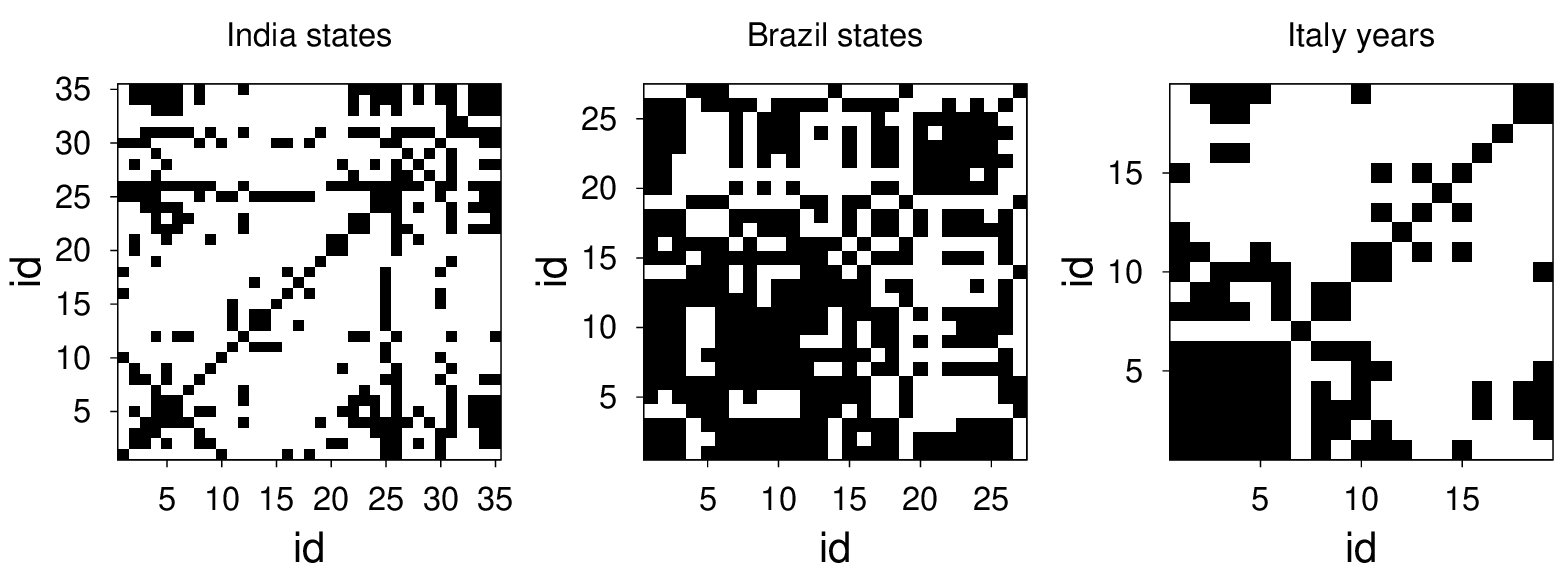}
\caption{Two sample Kolmorgorov-Smirnov test for pairs of data.
Left panel: pairwise test for all Indian states (2011-12, Id: state). Middle panel: pairwise test for all Brazilian states (2008-09, Id: state). Right panel: pairwise test for all Italian data across years (Id: year). Null hypothesis: two distributions are exactly identical. Colorcode: white square represents that the test rejects
the null hypothesis at 1\% level of significance. Black squares
represent the complementary scenario.}
\label{fig:KS_stat_all_countries}
\end{figure}

We do formal tests of how close the distributions are across states or across years, after
proper normalization. The results have been shown in Fig. \ref{fig:KS_stat_all_countries}.
It should be noted that the variable under consideration is
$\log(x/E(x))$ (Eqn. \ref{Eqn:scaling}). We show that for a substantial number of
cases, the normalized and log-transformed variables across states within a country (India and Brazil) or across time for the same country (Italy) are actually distributionally
identical. Thus not only the broad algebraic forms of the distributions (lognormal bulk and power law tail) coincide, but also the parameters describing the distributions are very similar.
In particular the Brazilian data establishes our claim. One interesting point
is that such tests could be sensitive to existence of fat tails. Given that both India and
Italy show prominent fat tails, it is not surprising that there are many cases where
the distributions are identical according to the test. Essentially this could be
attributed to the existence of outliers (data on extreme right) whereas the bulk
of the data do fall on a single distribution.

\section{An heterogeneous agent model} 
Here we propose a brief model of evolution of the consumption distribution. The 
basic goal would be to account for two observations, viz., the bulk of the 
distribution is seen to be following a lognormal distribution and there is a 
Pareto tail. Several assumptions are necessary to simplify the exposition. Time 
is discrete and goes till infinity i.e. $T=1,2,\ldots$. There are $N$ dynasties
who are producing and consuming. With a little abuse of notation, we will also 
use the same $N$ to denote the set of agents as well where no confusions arise. 
Each dynasty can be thought of as an unit of observation in the present context. 
We do not attempt to provide any microfoundation of their consumption decision 
and construct our model based on the approach recently 
introduced in \cite{Gabaix_11} and
\cite{Kelly_13}. They consider firm growth process resulting form 
interconnection among a large number of firms. The dynamical properties are 
developed from the proposed set of interconnections.
In this case, we follow a similar route and assume that the growth rate of 
consumption expenditure at the unit level (the unit could be individual or 
family or household depending on the case) 
admits interconnections between agents who differ in their attitude towards 
consumption. At the same time, we keep our model general enough to incorporate 
aggregate effects like long-term growth which can potentially affect inequality 
(positively or negatively). Hence, the growth rate of consumption expenditure is 
assumed to be a function of the level of present expenditure, household specific 
factors and the state of the macroeconomy.

In particular, we propose the following behavioral form of growth rate of 
expenditure of the $i$-th unit at any generic time-point $t$ 
\begin{eqnarray}
\hat{x}_i(t) &=& \frac{\Delta x_i(t)}{x_i(t-1)},  \nonumber\\
          &=& 
\lambda_i(t)\left(x_i(t-1)\right)^{\alpha_{i}(t)}+
\frac{\eta_i(t).m_{i}(t)+r(t)b_i(t)(\sum_j^Nw_j(t))+\chi_i(t)}{
x_i(t-1)}-1
\label{Eqn:main}
\end{eqnarray}
where $\alpha_i$ is agent specific shock, $\eta_i(t).m_i(t)$ is the total 
contribution of all unit-specific behavioral factors (e.g., religion, sex, geographic location etc.) that can 
potentially affect the level of consumption expenditure, $\chi_j(t)$
is a noise term with mean $\mu$ and variance $\sigma^2$. The term in the middle 
requires elaboration. We assume that consumption expenditure is affected at the 
business cycle frequency.
We are agnostic about the preferences of agents who 
participate in the same process and introduce a parameter $b_i(t)$ that captures 
the effects on the
$i$-th dynasty. The rate of return is given by $r(t)$ and the aggregate wealth 
(capital) is given by $\int w_j(t)dj$. 

There are some classic studies on the growth rates of consumption. In particular, 
\cite{Hall_78}
provided a framework to study the growth rates of consumption expenditure in the following form,
\begin{equation}
\hat{x}_i(t) =\frac{\chi_i(t)}{x_i(t-1)}.
\end{equation}
In App. \ref{Sec:app_rndwlk} we provide the basic framework that gives rise to such a growth rate. 
However, empirical studies have rejected such models. \cite{Jaeger_92} presents evidence that the theory is rejected when tested with U.S. data.
\cite{Haug_91} shows that the discrepancy might come from time aggregation bias (see also \cite{Molana_91}).
In a separate field of study, the firm growth rates had been described by similar functional forms. In particular,
\cite{Gabaix_11} starts describing a {\it granular} economy where each firm has a growth rate of
\begin{equation}
\hat{x}_i(t) =\chi_i(t)
\end{equation}
which is known to generate a lognormal distribution (see below). Such specifications are known as Gibrat's law (\cite{Gibrat:1931}). However, empirical estimations show that such a growth equation is 
incorrect (\cite{Hall_87}, \cite{Evans_87}, \cite{Calvo_06}).
\cite{Kelly_13} expands this simple framework to incorporate relationships between growth rates as follows,
\begin{equation}
\hat{x}_i(t)=f(\{\hat{x}_j(t)\}_j,W_{i,j\in N}(t),\chi(t) )
\end{equation}
where $f(.)$ captures a linear evolution of growth rates across size $x$ and $W$ captures the interaction matrix. 
We combine the above mechanisms to propose Eqn. \ref{Eqn:main}.

Imagine that the production function is 
given by the following simple equation
\begin{equation}
Y(t)=s(t)K(t)
\end{equation}
which says output is a linear function of capital ($K$) and a productivity shock 
$s$. One can incorporate labor. But for simplicity of exposition, we ignore it 
and assume that households (or individuals)
supply labor inelastically. The rental payment in the competitive market 
exhausts the total output. Noting that wealth acts as capital, we see that the 
income is given by 
$Y(t)=r(t)\int w_j(t)dj$ in absence of wage income. Note that this also introduces a coupling
among the agents through the capital market. This is the only source of direct interaction among agents.
Effectively such a factor induces an aggregate shock (rate or return $r$ is a function of the aggregate shock $s$) to the dynamic process. This type of approach to link the 
dynamic processes of multiple agents
can also be seen in \cite{SolomonGLV_98} although the links in the generalized Lotka-Volterra type models were considered for scaling purposes
(and not for representing any aggregate shock). 

We can decompose the income as the sum of a 
trend component ($Y^T(t)$) and a transitory component ($Y^C(t)$):
\begin{equation}
Y(t)=Y^T(t)+Y^C(t).
\end{equation}
The transitory component captures the purely fluctuating part.
Note that by construction
\begin{equation}
E(Y^C(t))=0.
\end{equation}
Since we have shown that there is an invariance in the distribution of 
expenditure once the data is normalized with respect to the household specific 
microeconomic factors, we ignore their contribution
for modeling the core inequality process. Also, the trend part can be ignored as 
it contributes to very long-term inequality. By taking all of the above into 
consideration, we arrive at the following equation,
\begin{equation}
\hat{x}_i(t) 
=\lambda_i(t)\left(x_i(t-1)\right)^{\alpha_{i}(t)}+\frac{\left(b_i(t)Y^C(t)+\chi_i(t)\right)}{x_i(t-1)}-1
\end{equation}
Therefore, we can use the definition of the growth rate and rewrite the above 
equation as
\begin{equation}
x_i(t) =\lambda_i(t)\left(x_i(t-1)\right)^{1+\alpha_{i}(t)}+b_i(t)Y^C(t)+\chi_i(t).
\label{Eqn:main_eqn}
\end{equation}

The business cycle component induces a distortion on the mechanism. Note that it is a common factor to all agents.
Thus if for a sustained period (a few quarters), the economy is either hit by very high or very low shocks expenditure
growth rate is affected exacerbating inequality. On the other hand when the economy returns to the baseline (zero shock),
the growth rates are diminished reducing inequality.
Business cycle can be taken to be an exogenous factor as usually it affects inequality and the converse is unlikely.
Thus due to the business cycle inequality might wax and wane (\cite{HeathcotePerri_15}).

Eqn. \ref{Eqn:main_eqn} forms the basis of the subsequent analysis.
Below we show known solutions of the above dynamic equation in three limits.

\begin{itemize}
\item[]{{\bf Case I.}} Let the transitory component and the agents' idiosyncratic shocks 
to expenditure be identically equal to zero i.e. $Y^C(t)=0$, $\alpha_i(t)=0$ and  $\lambda_i(t)=1$
for all $t$ in Eqn. \ref{Eqn:main_eqn}. Then the equation boils down to the 
following form 
\begin{equation}
\log(x_i(t)) = \log(x_i(t-1))+\chi_i(t).
\label{Eqn:unstable_lognorm}
\end{equation}
This random walk in logarithm is known to generate a lognormal distribution,
\begin{equation}
f(x,t)=\frac{1}{x(t)\sqrt{2\pi \sigma^2(t+1)}}\exp\left(-\frac{(\log 
x(t)-(t+1)\mu)^2}{2\sigma^2(t+1)} \right).
\end{equation}
Due to the explicit time dependence, there does not exist a steady state distribution for this process. Standard deviation 
increases without bound over time at the rate $\sqrt{t+1}$ (see for example \cite{chakrabarti2013econophysics}).

\item[]{{\bf Case II.}} Suppose the transitory component and the noise term are zero and the 
idiosyncratic term has a distribution over $[\alpha_{min},\alpha_{max}]$ where 
$0<1+\alpha<1$ for all $\alpha\in [\alpha_{min},\alpha_{max}]$ and all moments 
exist. Then we have 
\begin{equation}
\log(x_i(t)) =(1+\alpha_i(t)) \log(x_i(t-1))+\log \lambda_i(t).
\label{Eqn:stable_lognorm}
\end{equation}
By solving it recursively and using the lag operator $L$, we can rewrite it as
\begin{equation}
x_i(t) =\exp([1+(1+\alpha_i)L+(1+\alpha_i)^2L^2+\ldots].\lambda_i(t)).
\end{equation}
For simplicity of exposition we assumed $\alpha_i(t)=\alpha_i$ which can be relaxed without changing the basic result.
Thus the bracketed term becomes the sum of an infinite series of noise terms 
with standard deviation going to zero in limit. Hence, this process reaches a 
steady state described by a lognormal distribution. This process was formulated 
and proposed as a model of income evolution by Kalecki in 1945 (see 
\cite{chakrabarti2013econophysics} for a detailed exposition).

\item[]{{\bf Case III.}} Consider Eqn. \ref{Eqn:main_eqn} with the idiosyncratic 
term distributed over $[\alpha_{min},\alpha_{max}]$ where $E(\alpha)=0$ and 
$\sigma_{\alpha}\rightarrow 0$. We make two additional assumptions, (a) $E(\log \lambda_i(t))<0$ and
(b) $\eta_t=b_i(t)Y^C(t)+\chi_i(t)$, is distributed over $\mathbb{R}_+$.
\cite{SornetteCont_97} shows that under such assumptions, the steady state
distribution is power law. Such a dynamics is called
Kesten process
\begin{equation}
x(t)=\lambda(t)x(t-1)+\xi(t),
\end{equation}
which is known to generate power laws in the limit (\cite{Kesten_73}).
\cite{gabaix1999zipf1} uses such a mechanism to generate a power law in the city size distribution.
See also \cite{Sornette_book_06} for a textbook treatment.
\end{itemize}

\subsection{Heterogeneity of agents}
We introduce heterogeneity among the agents along one dimension viz., the upper range of the multiplicative factor ($\lambda_{max}$). Let us assume without loss of generalization that
$0< \lambda_{i, max}\le \lambda_{j, max}$ for all $i\le j$ and there exists some agent $1<k<N$ for whom $\lambda_{k,max}=0$. 
For all agents, $-1<\alpha_{min}\le \alpha_{max}\le 0$.
Thus effectively
there are two types of agents. Fraction 
$f$ of total number of agents have 
$0<\lambda_{i,max}<1$ for all $i\in N_{f}$ where
$N_f$ is the set of all such agents.
The evolution of the expenditure is given by Eqn. \ref{Eqn:main_eqn}, 
which we can rewrite as
\begin{equation}
\log(x_i(t)) =(1+\alpha_i(t)) \log(x_i(t-1))+\tilde{\lambda}_i(t)
\label{Eqn:agent-f}
\end{equation}
ignoring the noise factor and the business cycle 
variation. Thus we are back to Case II above.

The second type is described by $\lambda_{i,max}>1$ with the condition that $E(log(\lambda_i)<0)$ as mentioned in Case III above. To gain intuition about why
this process converges to a power law, assume that $E(\alpha)\rightarrow 0$ and $\sigma_{\alpha}\rightarrow 0$.
We assume that $b_i(t)$ is highly procyclical
i.e. $E(b_i.Y_C)>0$. The first assumption allows us to maintain the exact parametric 
requirement of the Kesten process. Strong procyclicality of consumption share effectively induces a lower bar on the expenditure even 
when 
the variable receives consecutive bad shocks through $\alpha_i$. Thus this 
becomes a reflective barrier and we can apply the methodology devised in the literature 
to find out the steady state distribution.

\begin{figure}
\begin{center}
\includegraphics[width=15cm]{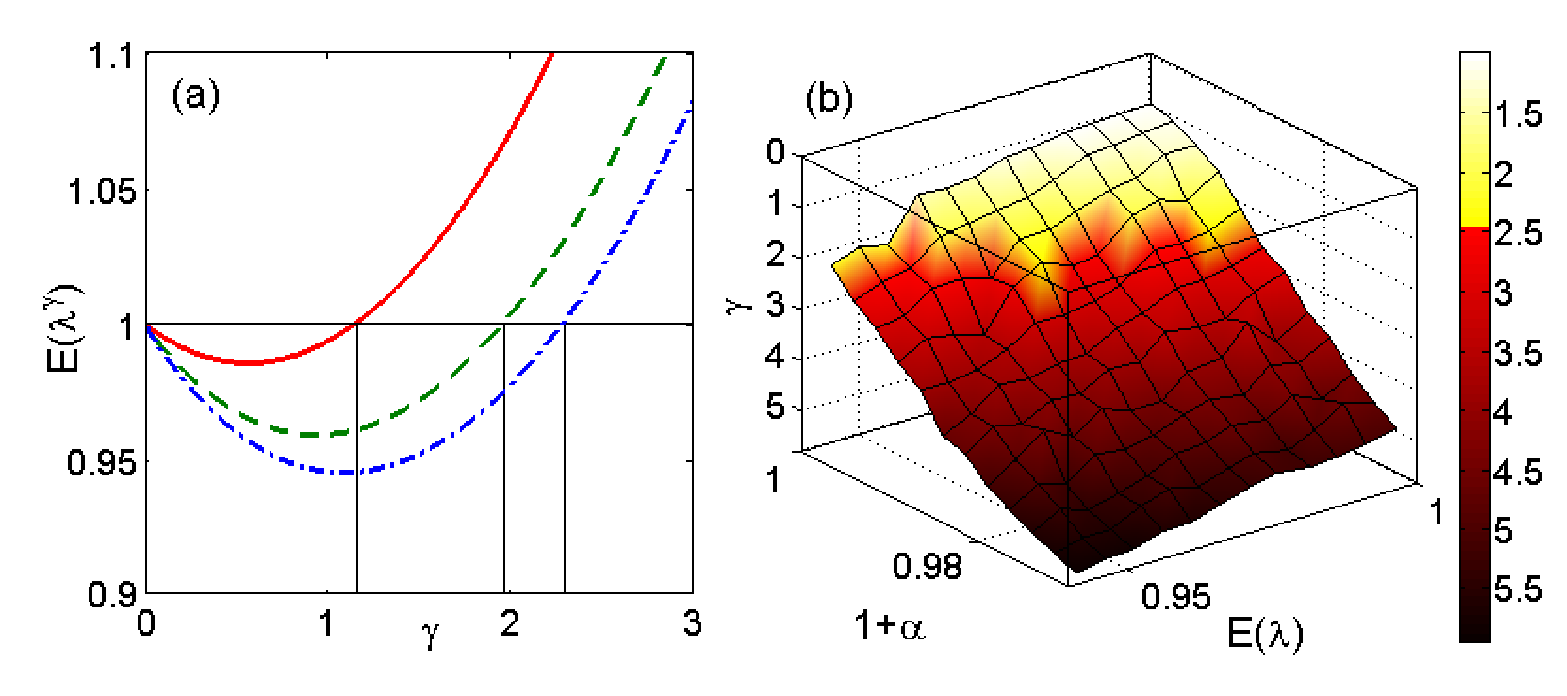}
\caption{Simulation results. Panel (a):  We plot the curve $E(\lambda^{\gamma})$
for 3 Monte Carlo realizations of the multiplicative shock $\lambda$ (with different averages) over a range of $\gamma$. At the points of intersection
with the horizontal line, one can find the theoretical prediction of the power law exponent $\gamma$ from Eqn. \ref{Eqn:power_law_condition}. Panel (b): Estimated Pareto exponent $\mu$ for simulations over ($\alpha,E(\lambda))$ parameter space. Colorbar shows the magnitude of $\gamma$ with a cut-off at 2.5 which is indicative of the empirically estimated coefficient of the consumption distribution for India.
}
 \label{fig:simulations}
\end{center}
\end{figure}

\cite{Gabaix_09} provides a very simple proof that the mechanism generates a power law. Assuming the existence of the lower (reflective) boundary through the business cycle effects, 
we know that
the variable can never be less than that. Hence, we consider the other extreme and study the right tail when the variable is far from the boundary making the additive terms relatively unimportant.
Let us assume that the multiplicative factor $\lambda_i$ is distributed according to 
$f(\lambda)$. Then we can write the evolution equation of the expenditure variable $X$ as
\begin{equation}
Prob.\left(X_i(t)<x\right)= Prob. \left(X_i(t)<\frac{x}{\lambda_i(t)}\right)
\end{equation}
Letting the left hand side be denoted as $M_{t}(e)$, we have a recursive equation
\begin{equation}
M_{t+1}(x)=\int_{\mathbb{R}_+}M_t\left(\frac{x}{\lambda}
\right)f(\lambda)d\lambda.
\end{equation}
The trick is to apply the criteria that when the system converges, the above equation
would be time independent and one can guess and verify the functional forms. In particular, \cite{Gabaix_09} shows that $M^{\infty}(X)\propto 1/X^{\gamma}$ solves the equation
and the condition reduces to
\begin{equation}
E(\lambda^{\gamma})=1.
\label{Eqn:power_law_condition}
\end{equation}
The same can also be shown using techniques developed by \cite{SornetteCont_97}.
Eqn. \ref{Eqn:power_law_condition} describes the relationship between the distribution
of the multiplicative factor and the exponent of the distribution.

In Fig. \ref{fig:simulations}, we present numerical simulations results for the evolution of the Pareto exponent
in the more general context (Eqn. \ref{Eqn:main_eqn}). Eqn. \ref{Eqn:power_law_condition}
gives the solution in only one limit ($\alpha \rightarrow 0$).  
The left panel shows the determination of the exponent following Eqn. \ref{Eqn:power_law_condition} in the limit. We use Monte Carlo simulation to find the Pareto exponent $\gamma$ in the general case. The right panel shows the estimated exponents
for $(\alpha,E(\lambda))$ pairs on the parameter plane. For the purpose of simulation, $\alpha$ is taken to be a constant. 
For each combination of $\alpha$ and $E(\lambda)$, we simulate
Eqn. \ref{Eqn:main_eqn} and estimate the exponent. The estimated exponents are 
averaged over $\mathcal{O}(10)$ realizations
in order to arrive at stable values. The surface indicates the exponents for different pairs of $\alpha$ and $E(\lambda)$.

\subsection{Shape of the ensemble distribution}
As we have described in the model above, there are essentially two types of agents. The first type generates a lognormal bulk
whereas the second type generates a power law distribution for the tail. Here we want to show that for the aggregate
distribution the tail is indeed described by a power law.

For this purpose, we use a result (see \cite{Gabaix_09} for a review) on the sum of two variables both distributed according power laws with potentially different exponents. 
Let the variables be $v_1$ and $v_2$. We assume that
\begin{equation}
x_i\sim C.x_i^{-(1+\gamma_i)}
\end{equation}
with $\gamma_1\ne \gamma_2$. Then the sum these two variables ($x=x_1+x_2$) will be distributed as
\begin{equation}
x\sim \bar{C}.x^{-(1+min(\gamma_1,\gamma_2))}.
\label{Eqn:power_theo}
\end{equation}
The intuition is simple: the fatter tail dominates the distribution. Note that the tail of a lognormal distribution 
can be approximated well by a power law with high exponent. Thus the tail of the aggregate distribution
can be modeled as the sum of two power laws with different exponents. Since the exponent of the distribution that
approximates a lognormal distribution is typically quite large, the other distribution dominates (following Eqn. \ref{Eqn:power_theo}).

\section{Discussion and summary}
In this paper, we describe two robust features of consumption inequality across time and countries. One, if consumption data is normalized with a proper scaling factor,
all data collapses on one single aggregate distribution. Two, the distribution has a lognormal bulk and a power law at the limit with high exponent (compared to income and wealth). Finally
we provide a stochastic model to account for the basic distributional features.

In the present work, we differentiate between long run versus short run inequality. We focus exclusively on the latter in order to
study cross sectional properties of inequality.   

Throughout our analysis, we have considered nominal data. There are two reasons for it. One, the available data is in nominal terms. Two,
in our cross-sectional analysis, we normalize the data first with respect to a scaling factor. As long as within region dispersion in price-levels are
not significantly high, such a normalization takes care of the pricing factors. All the subsequent analysis including cross-time and cross-country
comparisons are based on normalization with respect to respective scaling factor. Hence, such comparisons are free of biases due to between-region
or between-time periods variations in general price levels.

The way we have described the consumption growth process, a number of additional implications can be presented.
First, the power law arises due to the effects of
business cycle implying that inequality in cross-section can be affected at business cycle 
frequencies. \cite{HeathcotePerri_15} documents that in U.S. mean wealth is negatively correlated with macroeconomic volatility. One can argue that the changes in mean wealth is also accompanied by redistribution of purchasing power affecting inequality, thus corroborating the prior implication. \cite{Stiglitz_12} makes the point
that there is a relationship between fluctuations of macroeconomic fundamentals and inequality.
Secondly, in terms of the generative mechanism, our approach has a parallel with the method used by \cite{Benhabib_14} which also generates a power law in income in an overlapping generations framework. However, they
provided a microfounded framework for consumption-savings decision even though the essential mechanism is similar.
Earlier empirical works show that volatility of the business cycle is negatively related to the total
income of the country (\cite{Canning_98}). In terms of the model presented above, such a linkage would 
contribute to lower consumption volatility. In a fully specified utility-maximization framework this
would imply higher welfare. Finally, all other non-economic factors are seen to affect the mean of the expenditure distribution. This has a corollary that the spread of the core inequality process is independent of the
social, political and geographic factors. \cite{Angle_92} and \cite{Angle_93} also made a similar observation by considering U.S. data for specific social groups and conditioned on specific (e.g. racial or educational) factors. This
is complementary to our approach where 
we focus exclusively on the idea of core features of the distribution.

\section{Appendix} 
\label{Sec:app}
In this section, we present the additional figures and tables. A simple derivation of random walk model in consumption is also presented.

\subsection{Data summary}
\label{sec:app_data}
\begin{figure}
\includegraphics[width=15.5cm]{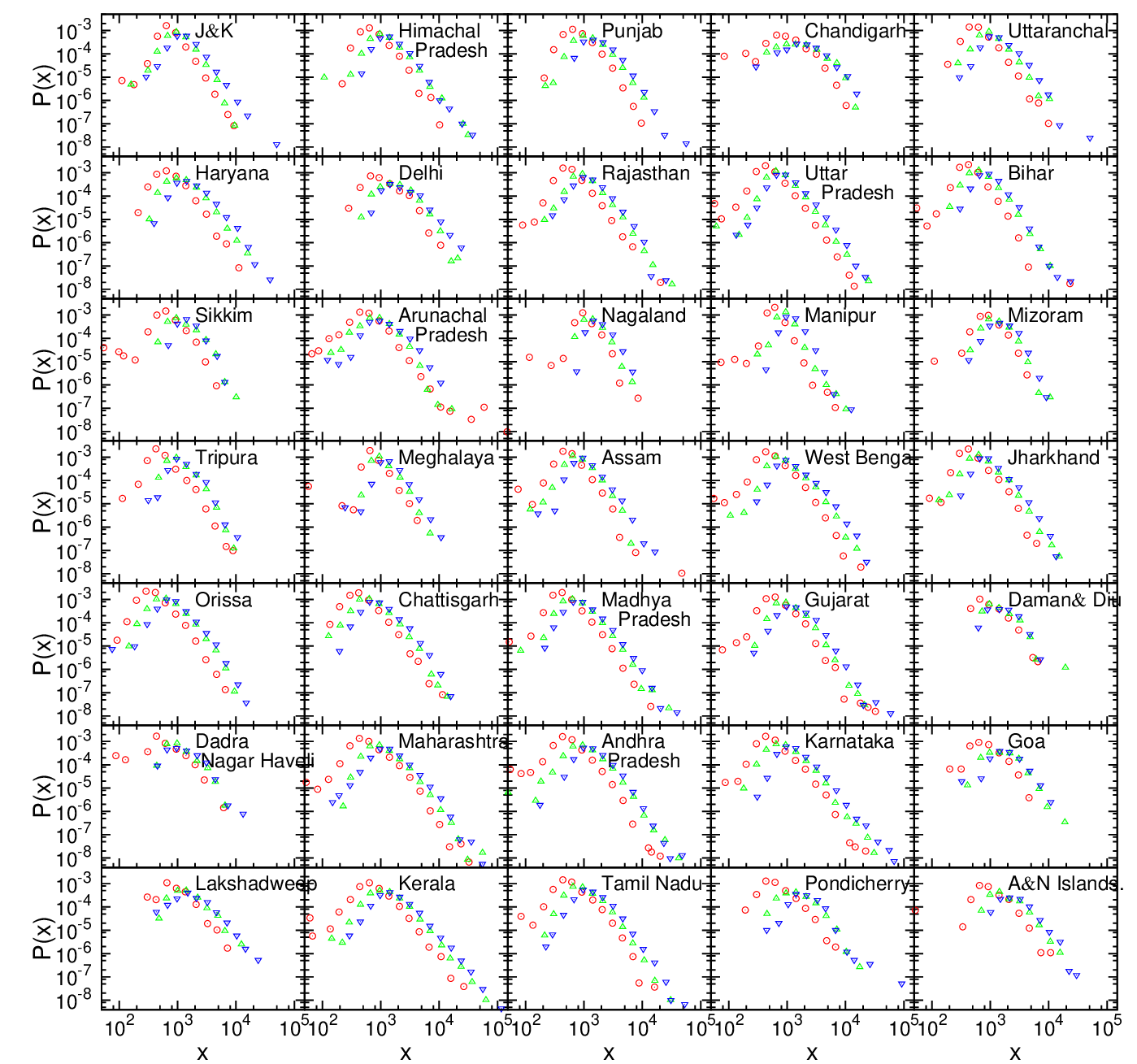}
\caption{Cross-sectional unnormalized data of Indian consumption expenditure 
across states has been shown
for 3 waves of data collection (red circles for 2004-05, 
green upward triangles for 2009-10, blue downward triangles for 2011-12).
Average growth in consumption over time is evident.
}
 \label{fig:Ind_allstates_3yrs}
\end{figure}

\begin{figure}
\includegraphics[width=15.0cm]{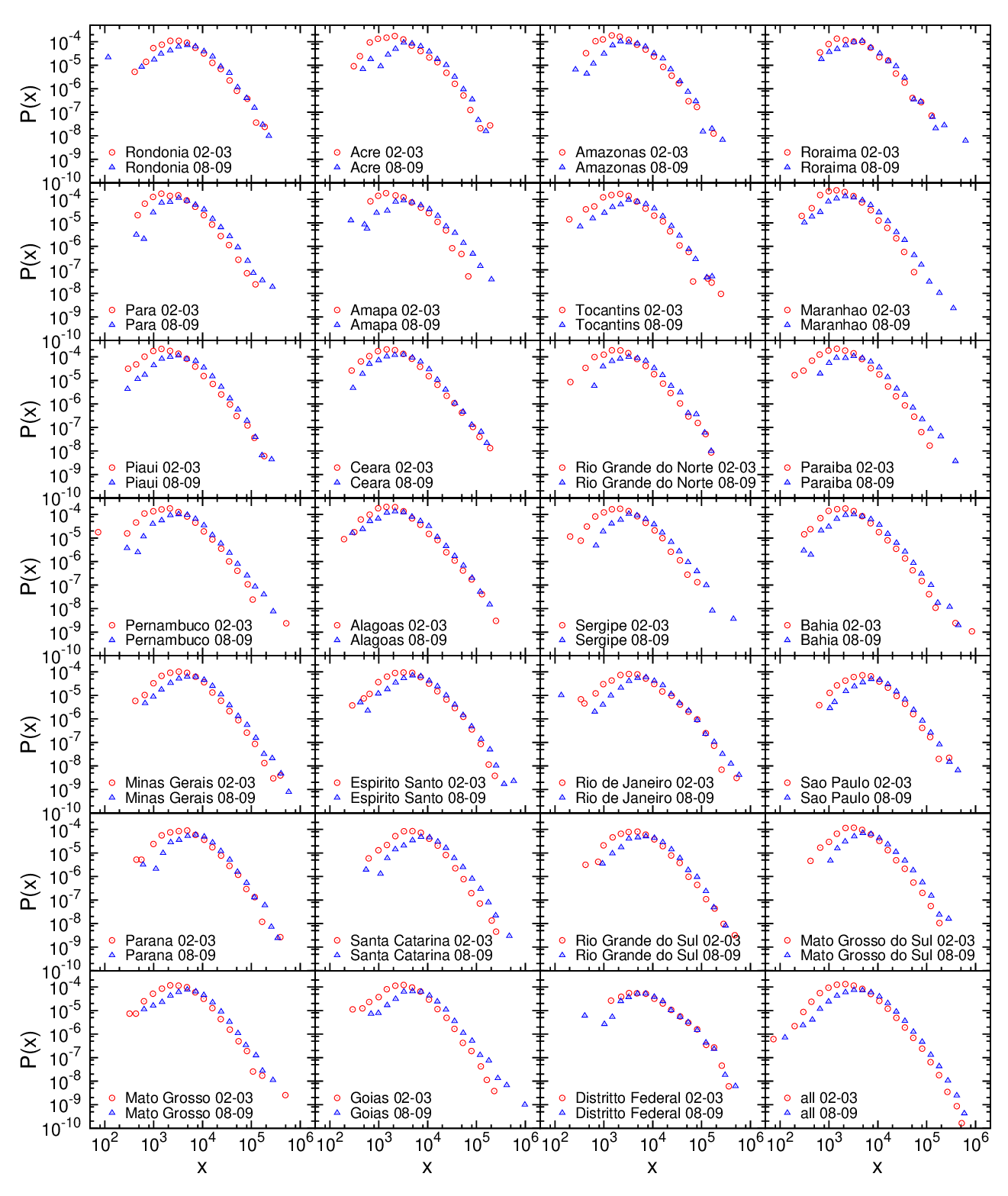}
\caption{Cross-sectional unnormalized data of consumption expenditure has been shown
for 2 waves of data collection (red circles for 2002-03, 
blue upward triangles for 2008-09).}
\label{fig:Brazil_allstates_2yrs}
\end{figure}

In Fig. \ref{fig:Ind_allstates_3yrs}, we present the available cross-sectional data
for India for all states for three waves of data collection (2004-05, 2009-10
and 2011-12). In the main text, we have analyzed data for 2011-12. \cite{CCGCN;16} 
presents some complementary results on the data set from 2009-10. Fig. \ref{fig:Ind_allstates_3yrs}
shows the general shift of the consumption density indicating both inflation and rising consumption power.
Fig. \ref{fig:Brazil_allstates_2yrs}
shows the available Brazilian data for all states for two waves of data collection
(2002-03 and 2008-09).

Details of the data and summary statistics have been tabulated
in table \ref{tab:brazil_states} (across states) and \ref{tab:brazil_urban} (urban-rural).
Table \ref{tab:italy} contains summary statistics for the Italian data
Finally, table \ref{tab:india68a} (across states) and \ref{tab:india68b} 
(social and other dimensions)
contains summary statistics for the Indian data.

\begin{table}[t]
\centering
\begin{tabular}{|l|c|l|c|c|c|c|c|c|}
\hline

ID & State &  & \multicolumn{3}{c|}{2002-2003}  & 
\multicolumn{3}{c|}{2008-2009}  \\ \cline{4-9}
&  index & & \#Households & $E(x)$  & Gini & \#Households & $E(x)$  & Gini \\
\hline 
1&11& Rond\^{o}nia &1112& 10870 & 0.535 & 907 & 15090.8 & 0.498\\ \hline  
2&12&Acre &960&8361.9& 0.570  & 863 & 12854.8 & 0.484 \\ \hline 
3&13& Amazonas &1075&7388.29& 0.549  & 1344 & 11011.6 & 0.504\\ \hline 
4&14& Roraima&554&9221.05& 0.529  & 644 & 12965.1 & 0.558\\ \hline 
5&15& Par\'{a} &1666&6645.31& 0.509  & 1894 & 12063.7 & 0.538\\ \hline 
6&16& Amap\'{a}&568&7163.86& 0.510  & 689 & 14428.5 & 0.537\\ \hline 
7&17& Tocantins&933&8027.05& 0.569  & 1270 & 12306.8 & 0.498\\ \hline 
8&21& Maranh\~{a}o&2231&4749.73& 0.502 &2562  & 8725.61 & 0.524\\ \hline 
9&22& Piau\'{i}&2222&6263.12& 0.557 & 2056 & 9844.81 & 0.498\\ \hline 
10&23& Cear\'{a}&2017&6351.92& 0.571 & 1861 & 8553.52 & 0.514\\ \hline 
11&24& Rio Grande do Norte&1548&6819.38& 0.558  & 1342  &11016.7& 0.501\\\hline 
12&25& Para\'{i}ba&2367&5614.04& 0.538  &  1628 & 10990.6 & 0.543\\ \hline 
13&26& Pernambuco&1674&7126.29& 0.558  &  2367 & 11360.2 & 0.532 \\ \hline 
14&27& Alagoas&2965&6601& 0.583 &  2712  & 9138.34 & 0.541\\ \hline 
15&28& Sergipe&1143&6705.08& 0.518  & 1654 & 12289.4 & 0.512\\ \hline 
16&29& Bahia&2457&7678.18& 0.584 &  3050 & 11997  & 0.539\\ \hline 
17&31& Minas Gerais&3004&11283.4& 0.528 &  5028 & 17065.8 & 0.508\\ \hline 
18&32& Esp\'{i}rito Santo&2337&12148.5& 0.535 &  3489 & 16538.2 & 0.511\\ 
\hline 
19&33& Rio de Janeiro&1285&17973.1& 0.591  &  1938  & 22063.4 & 0.551\\ \hline 
20&35& S\~{a}o Paulo&2017&16074.3& 0.516  &  3623 & 22592.2  & 0.486\\ \hline 
21&41& Paran\'{a}&2263&12733.5&  0.519  &  2477 & 17829.8 & 0.470\\ \hline 
22&42& Santa Caterina&1989&12169.9& 0.464  &  2029 & 23447.8 & 0.498\\ \hline 
23&43& Rio Grande do Sul&1850& 14370.6& 0.534  &  2210 & 20394.7 & 0.482\\ 
\hline 
24&50& Mato Grosso do Sul&2541& 9788.09& 0.505  &  2247 & 17071.5 & 0.498\\ 
\hline 
25&51& Mato Grosso&2355&9241.06& 0.513  &  2423 & 14363.5 & 0.488\\ \hline 
26&52& Go\'{i}as&2356&9160.95& 0.505  & 2686 & 16997.5  & 0.523\\ \hline 
27&53& Distrito Federal&981&26497.2& 0.590  &  977 & 26081.3 & 0.564\\ \hline
& & all Brazil & 48470 &9626.94& 0.568 & 55970 & 12777& 0.533\\ \hline 
\end{tabular}
\caption{Number of households, average per capita consumption expenditure 
$E(x)$ and Gini indices for Brazil, for 2 rounds 2002-2003 and 2008-2009.
Tabulted accross states.}
\label{tab:brazil_states}
\end{table}

\begin{table}[t]
\centering
\begin{tabular}{|l|c|c|c|c|c|c|}
\hline
 Location & \multicolumn{3}{c|}{2002-2003}  & 
\multicolumn{3}{c|}{2008-2009}  \\ \cline{2-7}
  & \#Households & $E(x)$  & Gini & \#Households & $E(x)$  & Gini \\
\hline 
 rural & 48357 &18352.6& 0.514  & 43193 & 16658.7 & 0.528 \\ \hline 
 urban& 114 & 20352.3 & 0.478  & 12777 & 9847.44 & 0.507\\ \hline  
\end{tabular}
\caption{Number of households, average per capita consumption expenditure 
$E(x)$ and Gini indices for Brazil, for 2 rounds 2002-2003 and 2008-2009. Tabulted according to location.}
\label{tab:brazil_urban}
\end{table}

\begin{table}[t]
\centering
\begin{tabular}{|l|c|c|c|c|}
\hline

ID & year &  \#Households & $E(x)$  & Gini \\\hline 
1&1980&2980 &8529.86 & 0.307 \\ \hline  
2&1981&4091 &10373.1 & 0.298\\ \hline  
3&1982&3967 &12304.9 & 0.296\\ \hline  
4&1983&4107 &13952.4 & 0.296 \\ \hline  
5&1984&4172 &15474.4 & 0.302\\ \hline  
6&1986&8022 &17315.6 & 0.297\\ \hline  
7&1987&8024 &23652.2 & 0.333\\ \hline  
8&1989&8274 &24392.9 & 0.289\\ \hline  
9&1991&8186 &26334.6 & 0.285\\ \hline  
10&1993&8088 &29087.4 & 0.297\\ \hline  
11&1995&8135 &33631.8 & 0.305\\ \hline  
12&1998&7146 &36157.1 & 0.316\\ \hline
13&2000&8001 &38089.7 & 0.308\\ \hline  
14&2002&8010 &20466.6 & 0.317\\ \hline  
15&2004&8011 &22419.9 & 0.305\\ \hline  
16&2006&7768 &23674.8 &  0.290\\ \hline  
17&2008&7976 &23817.4 & 0.279\\ \hline  
18&2010&7950 &25261.1 & 0.294\\ \hline
19&2012&8149 &25408.3 & 0.291\\ \hline
\end{tabular}
\caption{Number of households, average per capita consumption expenditure 
$E(x)$ and Gini indices for Italy, for several years.}
\label{tab:italy}
\end{table}

\begin{table}[t]
\centering
\begin{tabular}{|l|l|c|c|c|c|}
\hline
ID& Geographic location &  \#Households & $E(x)$  & Gini\\\hline 
1&Jammu \& Kashmir	& 3382 & 1846.749 &  0.310  \\ \hline
2&Himachal Pradesh	& 2040 & 2105.473 &  0.336  \\ \hline
3&Punjab	& 3118 & 2571.475 &  0.334  \\ \hline
4&Chandigarh	& 312 & 3577.070 &  0.378  \\ \hline
5&Uttaranchal	& 1784 & 2073.443 &  0.350  \\ \hline
6&Haryana	& 2589 & 2575.453 &  0.365  \\ \hline
7&Delhi	& 999 & 3653.659 &  0.382  \\ \hline
8&Rajasthan	& 4127 & 1824.600 &  0.332  \\ \hline
9&Uttar Pradesh	& 9018 & 1414.226 &  0.357  \\ \hline
10&Bihar	& 4581 & 1243.082 &  0.286  \\ \hline
11&Sikkim	& 768 & 1850.008 &  0.243  \\ \hline
12&Arunachal Pradesh	& 1674 & 1863.248 &  0.371  \\ \hline
13&Nagaland	& 1024 & 2185.466 &  0.241  \\ \hline
14&Manipur	& 2560 & 1438.841 &  0.220  \\ \hline
15&Mizoram	& 1536 & 2129.177 &  0.259  \\ \hline
16&Tripura	& 1856 & 1609.395 &  0.290  \\ \hline
17&Meghalaya	& 1260 & 1759.212 &  0.263  \\ \hline
18&Assam	& 3440 & 1417.833 &  0.309  \\ \hline
19&West Bengal	& 6317 & 1886.182 &  0.387  \\ \hline
20&Jharkhand	& 2737 & 1349.507 &  0.341  \\ \hline
21&Orissa	& 4029 & 1246.751 &  0.347  \\ \hline
22&Chattisgarh	& 2173 & 1464.659 &  0.367  \\ \hline
23&Madhya Pradesh	& 4718 & 1449.213 &  0.366  \\ \hline
24&Gujarat	& 3430 & 2143.533 &  0.345  \\ \hline
25&Daman \& Diu	& 128 & 2196.510 &  0.273  \\ \hline
26&Dadra \& Nagar Haveli	& 192 & 1901.413 &  0.335  \\ \hline
27&Maharashtra	& 8041 & 2323.568 &  0.391  \\ \hline
28&Andhra Pradesh	& 6898 & 2094.464 &  0.345  \\ \hline
29&Karnataka	& 4096 & 2117.983 &  0.399  \\ \hline
30&Goa	& 448 & 2700.791 &  0.306  \\ \hline
31&Lakshadweep	& 192 & 3094.633 &  0.396  \\ \hline
32&Kerala	& 4460 & 3014.732 &  0.431  \\ \hline
33&Tamil Nadu	& 6647 & 2122.480 &  0.357  \\ \hline
34&Pondicherry	& 576 & 3086.998 &  0.339  \\ \hline
35& Andaman \& Nicobar Is. & 567 & 3937.967 &  0.347  \\ \hline
& all India	& 101717 & 1939.779 &  0.378  \\ \hline
\end{tabular}
\caption{Number of households, average per capita consumption expenditure 
$E(x)$, Gini index for India. Data available for different 
states for 68st round (2011-2012).}
\label{tab:india68a}
\end{table}

\begin{table}[t]
\centering
\begin{tabular}{|l|c|c|c|c|}
\hline
Filter &  \#Households & $E(x)$  & Gini \\\hline 
ST	& 13403 & 1601.763 &  0.338  \\ \hline
SC	& 15652 & 1507.782 &  0.335  \\ \hline
OBC	& 39721 & 1800.488 &  0.360  \\ \hline
Other castes	& 32938 & 2450.539 &  0.391  \\ \hline \hline 
Hinduism	& 77036 & 1935.365 &  0.384  \\ \hline
Islam	& 13274 & 1698.742 &  0.347  \\ \hline
Christianity	& 6930 & 2223.477 &  0.357  \\ \hline
other religions	& 4477 & 2291.247 &  0.359  \\ \hline \hline 
Rural	& 59693 & 1525.498 &  0.322  \\ \hline
Urban	& 42024 & 2528.244 &  0.386  \\ \hline
\end{tabular}
\caption{Number of households, average per capita consumption expenditure 
$E(x)$, Gini index for India. Data available  religions, 
caste as well as urban-rural divide for 68th round (2011-2012).}
\label{tab:india68b}
\end{table}

\subsection{Random walk in consumption}
\label{Sec:app_rndwlk}
In a standard utility-maximizing framework with representative agent, the Euler equation would be
\begin{equation}
u'(x_t)=\beta R(t)E_t(u'(x_{t+1}))
\end{equation}
where $u(.)$ is the utility function defined over consumption good $x$ at time $t$. The discount factor is denoted by $\beta$ and the rate of return by $R$. $E_s(.)$ denotes expectation with the information set $s$. The simplest framework to derive random walk (\cite{Hall_78}) is to assume
\begin{equation}
u(x)=-\frac{(\bar{x}-x)^2}{2}
\end{equation}
where $\bar{x}$ is the bliss point. Also assume $R\beta=1$ to solve the above equation to get
\begin{equation}
E_t(x_{t+1})=x_t.
\end{equation}
Thus the consumption growth equation is 
\begin{equation}
x_{t+1}=x_t+\chi_{t+1},
\end{equation}
where $\chi_{t+1}$ is the innovation term. Thus the growth rate is
\begin{equation}
\hat{x}(t)=\frac{\chi (t)}{x(t-1)}.
\end{equation}


\end{document}